\documentstyle[twocolumn,prb,epsfig,aps]{revtex}
\begin{document}
\draft
\title{Magnetoresistivity and Complete $H_{c2}(T)$ in $MgB_2$.}
\author{S. L. Bud'ko, C. Petrovic, G. Lapertot\thanks {On 
leave from Commissariat a l'Energie Atomique, DRFMC-SPSMS, 38054 Grenoble,
France},
C. E. Cunningham\thanks{On leave from Dept. of Physics, 
Grinnell College, Grinnell, IA 50112}, and P. C. Canfield}
\address{Ames Laboratory, U.S. Department of Energy and Department of Physics and Astronomy\\
Iowa State University, Ames, Iowa 50011}
\author{M-H. Jung\thanks{also at Physics Dept., New Mexico State University, 
Las Cruces, New Mexico} and A. H. Lacerda}
\address{National High Magnetic Field Laboratory - Pulse Facility, 
Los Alamos National Laboratory, MS E536 Los Alamos, NM 87545}

\date{\today}
\maketitle
\begin{abstract}
Detailed magneto-transport data on dense wires of $MgB_2$ are reported for
applied
magnetic fields up to 18 T. The temperature and field dependencies of the 
electrical resistivity are consistent with $MgB_2$ behaving like
a simple metal and
following a generalized form of Kohler's rule. In addition, given the
generally
high $T_c$ values and narrow resistive transition widths associated 
with $MgB_2$ synthesized in this manner, combined with applied magnetic 
fields of up to 18 T, an accurate and complete $H_{c2}(T)$ curve could be 
determined. This curve agrees well with curves determined from lower 
field measurements on sintered pellets and wires of $MgB_2$. $H_{c2}(T)$ 
is linear in $T$ over a wide range of temperature (7 K $\le~T~\le$ 32 K) and 
has an upward curvature for $T$ close to $T_c$. These features are similar 
to other high $\kappa$, clean limit, boron-bearing intermetallics: 
$YNi_2B_2C$ and $LuNi_2B_2C$.
\end{abstract}


\section{Introduction}

The recent discovery\cite {jap} of superconductivity in $MgB_2$ has lead to
a flurry of activity.
Measurements of the boron isotope effect\cite {budko} show that there is
a shift in $T_c$ from
39.2 K to 40.2 K (for $Mg^{11}B_2$ and $Mg^{10}B_2$ respectively),
a result consistent with electron phonon mediated BCS
superconductivity. Measurements\cite {budko,DKF,canfield,jap1,wisc,caplan} of
the upper critical field, $H_{c2}(T)$, and
the thermodynamic critical field, $H_c(T)$, as well as the specific 
heat are 
all consistent with $MgB_2$ being a fairly typical intermetallic
superconductor
with an atypically high transition temperature. Although single 
crystal samples are not yet available, it has recently been found that 
dense, very high quality, wire samples of $MgB_2$ can be
made.\cite {canfield} These samples
have a superconducting transition temperature above that found for
$Mg^{11}B_2$,
as would be expected based on the natural abundance of $^{10}B$. These
samples allow
for the direct measurement of electrical resistivity and, given that the width
of the superconducting transition is narrower in $MgB_2$ wire than in powder 
or sintered pellet samples,\cite {DKF,canfield} wire samples allow for an 
accurate determination 
of the upper critical field $H_{c2}(T)$.  In this communication we present data on 
the magneto-transport of $MgB_2$ wires for applied magnetic fields of up to 
18 T and over the temperature range 1.5 - 300 K. By a careful analysis of 
the resistivity data we are able to conclude that $MgB_2$ behaves like
a simple metal
in the normal state with all of our magnetoresistance data collapsing onto a 
single curve in accordance to Kohler's rule. In addition we are able to 
construct the full $H_{c2}(T)$ curve. We find that $H_{c2}(T)$ is linear 
over a much larger temperature range than would be expected,\cite {WHH}
leading to a $H_{c2}(0)~\approx$ 16.4 T.

\section{Experimental methods}

$MgB_2$ wire was produced\cite {canfield} by sealing 100 $\mu$m diameter boron fiber
and $Mg$ into a $Ta$ tube with a nominal ratio of $Mg_2B$.  Given that $MgB_2$ is 
the most $Mg$ rich 
binary $Mg-B$ compound known, it was felt that excess $Mg$ would aid in the 
formation of the proper, stoichiometric phase.  The sealed $Ta$ tube was itself
sealed
in quartz and then placed into a 950$^\circ$C box furnace for two 
hours. The reaction ampoule was then removed from the furnace and quenched to room 
temperature.  
Whereas the boron fiber has a diameter of 100 $\mu$m, the $MgB_2$ wire has
a diameter of
approximately 160 $\mu$m.  Although the $MgB_2$ wires are somewhat brittle,
the integrity
of the filament segments is preserved during the exposure to the $Mg$ vapor; 
i.e. the fibers did not decompose, fragment, or turn into powder.  The
resulting
wire has over 80\% the theoretical density of $MgB_2$ and measurements of the
temperature
dependent resistivity reveal that $MgB_2$ is highly conducting in the normal
state.
The room temperature resistivity has a value of 9.6 $\mu$Ohm-cm; whereas the 
resistivity at $T$ = 40 K is 0.38 $\mu$Ohm-cm.  The zero field $T_c$ value for the 
wire sample is higher than that found for $Mg^{11}B_2$, a result consistent 
with the natural abundance of $^{10}B$.  It should be noted that both wire 
and sintered pellet samples synthesized in this manner tend to have very 
high and sharp superconducting transitions.

Magnetoresistivity measurements utilizing a 20 T, superconducting magnet
were performed
at the National High Magnetic Field Laboratory, Pulsed Facility. 
A standard four-probe ac method was used, utilizing Epotek H20E silver epoxy for making 
electrical contacts. The contact resistance was approximately 1 Ohm.  Given the 
well-defined geometry of the samples, accurate  
measurements of resistivity were possible. The ac current was applied along the 
wire and the magnetic field was applied perpendicular to the current direction. 
The sample was mounted in a flow cryostat able to regulate the temperature 
from 1.4 K to room temperature. 

\section{Data and analysis}
Figure 1 presents temperature dependent electrical resistivity data for 
a $MgB_2$ wire sample taken at a variety of applied fields for $H~\le$ 18 T. 
Two features are clearly seen: there is a suppression of the superconducting 
phase to lower temperatures for increasing applied field, and there is a clear, 
large magnetoresistivity in the normal state.  Looking first at the suppression 
of superconductivity, the inset to Fig. 1 presents an enlarged
view of the low
temperature resistivity data. Using these data, three temperatures can be 
extracted from each curve: onset temperature, temperature of maximum $d\rho/dT$, 
and completion temperature, where  onset and completion temperatures are 
determined by extending the maximum $d\rho/dT$ line up to the normal state 
and down to zero resistivity.  

Figure 2 presents the $H_{c2}(T)$ curve that we deduce from these data.  
In addition to the high field data taken at NHMFL (shown as open symbols), 
data taken in a Quantum Design PPMS system at lower fields on a sample 
from the same batch are also shown (filled symbols).  Several features of this 
curve are worth noting.  First of all it has a large temperature / field range 
over which it is linear (7 K $\le T \le$ 32 K).  Below approximately 7 K 
$H_{c2}(T)$ starts to roll over and saturate.  This leads to a low temperature 
value of $H_{c2}$(1.5 K) = 16.2 T, which is significantly larger than 
estimates\cite {DKF} based on the assumption\cite {WHH} that the low
temperature
$H_{c2}$(0) = 0.71 $T_c~[dH_{c2}(T)/dT]$ = 12.5 T.  Secondly, at high 
temperatures ($T~\ge$ 32 K) there is a distinct positive, upward curvature 
associated with the $H_{c2}(T)$ curve.  This is not unique to the current 
form of our sample but was also seen in resistively determined $H_{c2}(T)$ 
for sintered pellets of $Mg^{10}B_2$.\cite {DKF}  Taken as
a whole, the temperature
dependence of $H_{c2}$ for $MgB_2$ is remarkably similar to that recently 
found for other non-magnetic, intermetallic, boride superconductors:  
$LuNi_2B_2C$ and $YNi_2B_2C$.\cite {boro1,boro2,boro3}  In these
cases $H_{c2}(T)$ is linear over
an extended region of $T$ and near $T_c$ there is a distinct upward curvature.  
In both the case of $MgB_2$ as well as in the case of $Y/LuNi_2B_2C$ the 
material is a high $\kappa$, type-II superconductor and in both cases 
the as grown compounds are well within the clean
limit.\cite {DKF,canfield,boro1,boro2,boro3}

Turning to the normal state magnetoresistivity, Fig. 3 shows 
$\Delta\rho/\rho_0$ vs $H/\rho_0$ on a $log-log$ plot to demonstrate that all of the 
data presented in Fig. 1, as well as the two isothermal $\rho(H)$ plots shown in the 
inset of Fig. 3,  are broadly consistent with the generalized form of Kohler's rule.  
The fact that all of these data fall (roughly) onto a single curve implies that there 
is a single salient scattering time in the normal state transport
of $MgB_2$.\cite {pippard}
This is what would be anticipated for a simple nonmagnetic intermetallic
sample.  It is worth
noting that such a clear magnetoresistance would be much harder to
detect in 
samples with enhanced impurity or defect scattering.\cite{korea}

The isothermal $\Delta\rho(H)/\rho_0$ data shown in the inset of Fig. 3 can
be fit
to $\Delta\rho(H)/\rho_0~\propto~H^{\alpha}$ with $\alpha$ = 1.4-1.5.  
In addition, the temperature dependent normal state resistance of sintered 
$Mg^{10}B_2$ pellets as well as the resistivity of wire samples in zero field
can be fit to a power law between $T^{2.6}$ and $T^3$ at low temperatures
($T~\le$ 200 K).\cite {DKF,canfield}

Finally, there is a slight upturn observed in the low-temperature
$\rho(T)$ data taken
in high applied magnetic fields (Figure 1). Although the origin of this
feature is not completely understood, similar features have been
observed for other high purity, nonmagnetic intermetallic
compounds.\cite {dia1,dia2}

\section{Conclusions}

In this communication we present detailed magetoresistivity data on dense
(over 80\%), high quality
samples of $MgB_2$.  In the normal state we find that $MgB_2$ has a temperature and 
field dependent resistivity that is consistent with $MgB_2$ being a highly
conducting
intermetallic compound that can be synthesized so as to achieve very low 
residual resistivities.  This allows the large values of
$\Delta\rho/\rho_0$ to
reveal themselves.  The fact that the magnetoresistivity data follow
Kohler's rule
is consistent with $MgB_2$ behaving like a simple metal with one dominant
scattering time.\cite{pippard}

In the superconducting state we find that $MgB_2$ has a relatively linear 
$H_{c2}(T)$ curve with slight deviations from linearity at both high and low 
temperatures.  This leads to a relatively high value of 
$H_{c2}$(0) = 16.4 T.  The linear behavior of $H_{c2}(T)$ as well as the 
upward curvature in $H_{c2}(T)$ for $T$ near $T_c$ is similar to behavior seen 
in other boron bearing intermetallic compounds ($YNi_2B_2C$ and $LuNi_2B_2C$) 
that have large $\kappa$ values and are within the clean limit.  
This similarity brings up the obvious question of whether such temperature 
dependencies of $H_{c2}(T)$ are a generic feature associate with this 
subclass of intermetallic superconductors.

\section{Acknowledgments}

We would like to thank D. K. Finnemore for useful discussions.
Ames Laboratory is operated for the U. S. Department of Energy by Iowa State University 
under Contract No. W-7405-Eng.-82.  This work was supported by the director for Energy 
Research, Office of Basic Energy Sciences.  Work performed at the National High Magnetic 
Field Laboratory was supported by the National Science Foundation, 
The State of Florida and the U. S. Department of Energy. One of us 
(M-H J) acknowledges partial support form LANSCE - LANL.


\begin{figure}
\epsfxsize=0.9\hsize
\vbox{
\centerline{
\epsffile{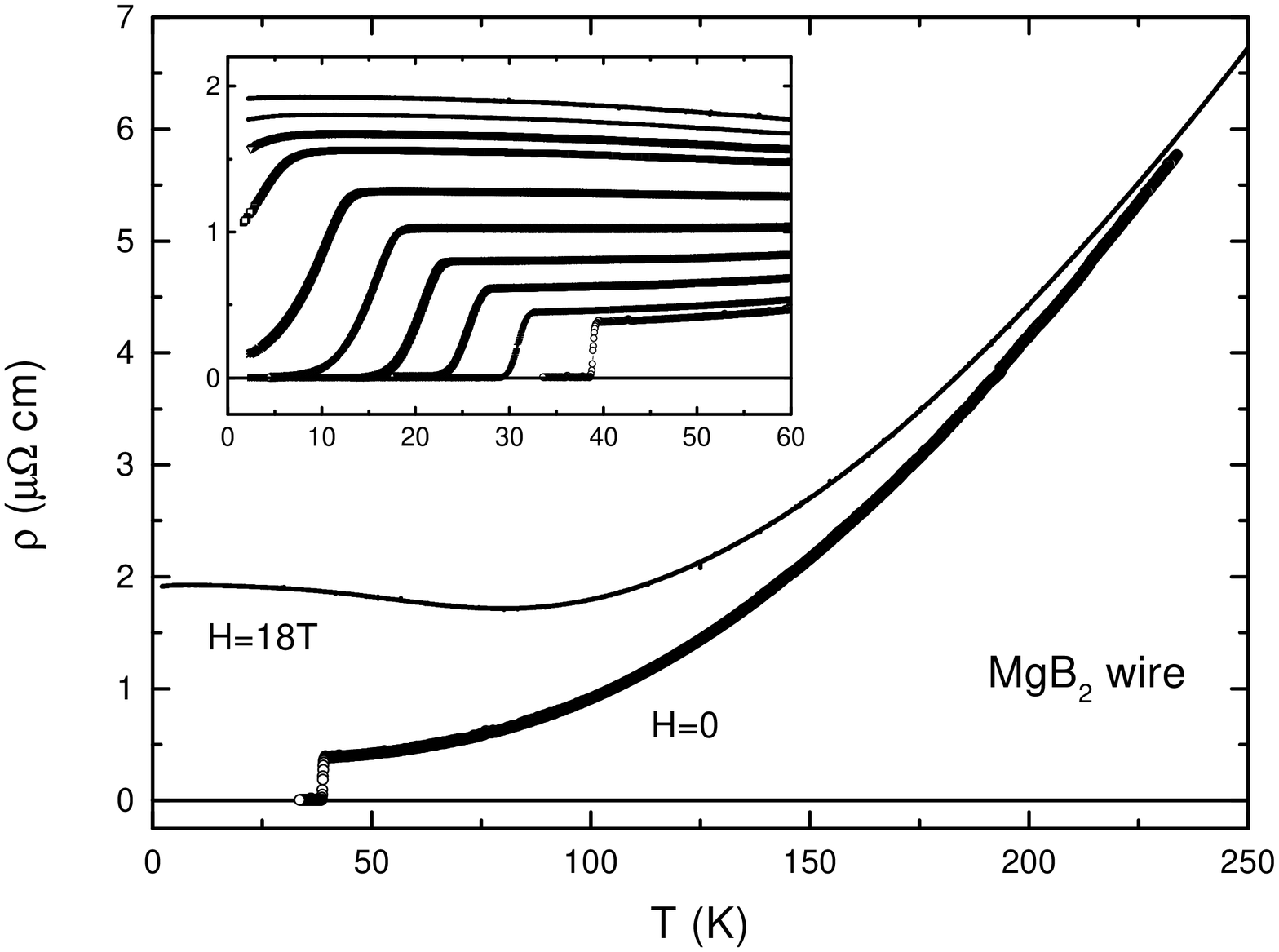}
}}
\caption{Temperature dependent resistivity of $MgB_2$ wire in zero and 18 T applied 
field. Inset: low temperature $\rho(T)$ data in applied fields of (from bottom right 
to top left) 0, 2.5, 5, 7.5, 10, 12.5, 15, 16, 17, 18 Tesla.} 
\label{F1}
\end{figure}
\begin{figure}
\epsfxsize=0.9\hsize
\vbox{
\centerline{
\epsffile{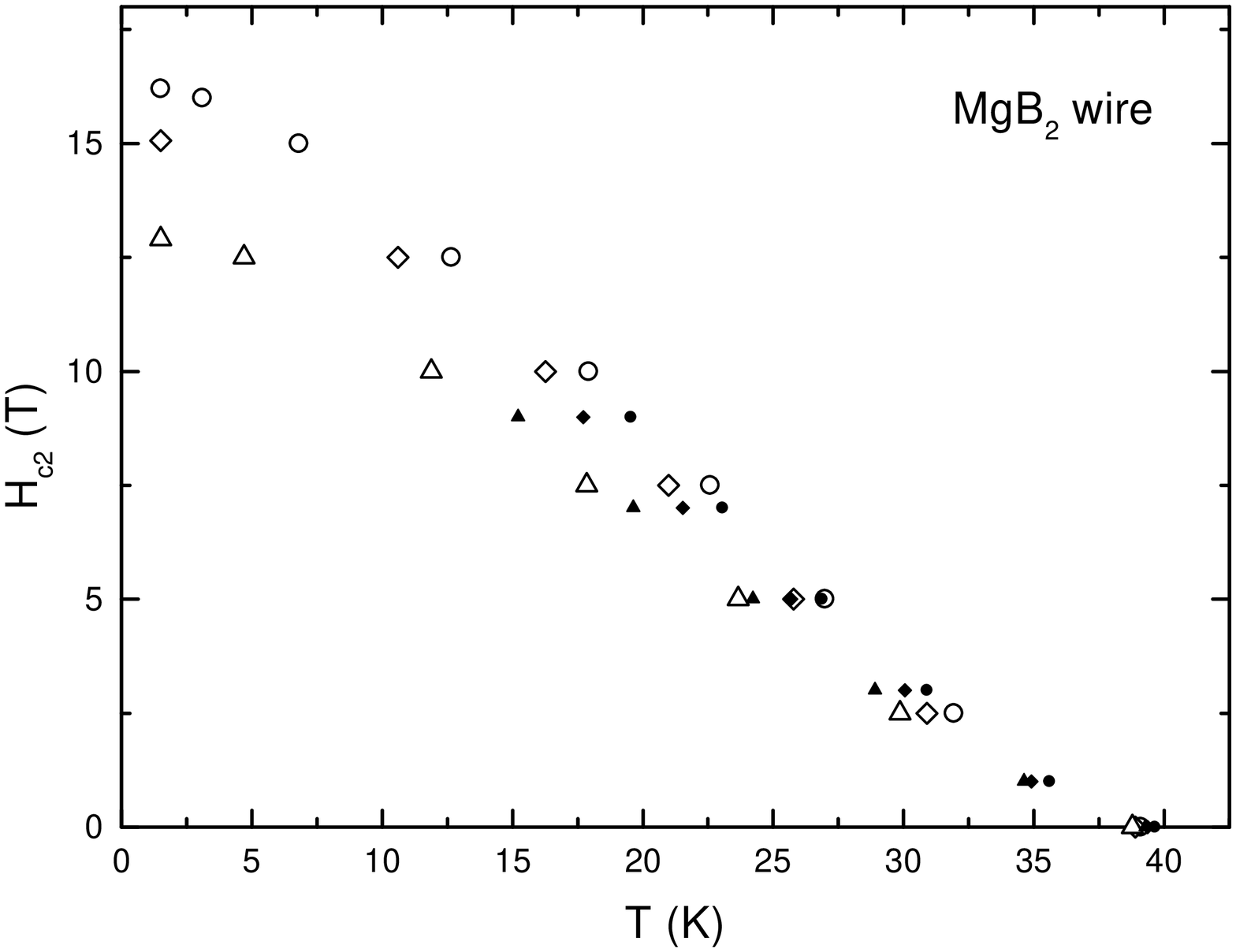}
}}
\caption{Upper critical field of $MgB_2$ wires. Different symbols represent different 
criteria (onset, maximum of $d\rho/dT$ and completion) as described in text.
Data at 1.5 K are
from $\rho(H)$ measurements (not shown). Open symbols - this work
(Fig. 1), filled symbols - from Ref. [4].}
\label{F2}
\end{figure}
\begin{figure}
\epsfxsize=0.9\hsize
\vbox{
\centerline{
\epsffile{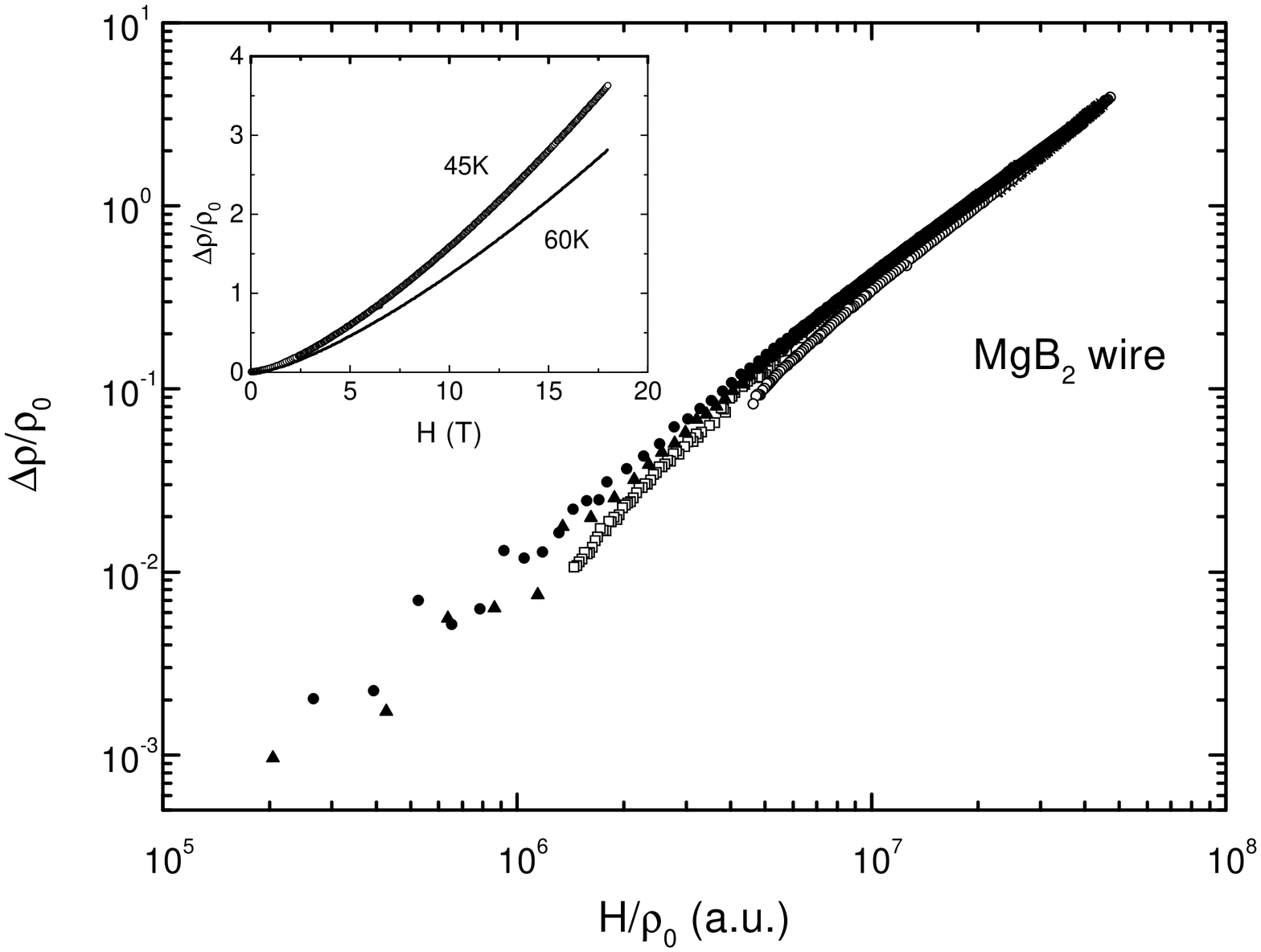}
}}
\caption{Kohler's plot for $MgB_2$ wires: open symbols from temperature-dependent 
resistivity; filled symbols from field-dependent resistivity at 45 K and 60 K. 
Inset $\Delta\rho(H)/\rho_0$ at 45 K and 60 K.}
\label{F3}
\end{figure}

\vfil\eject


\begin{references}  
\bibitem{jap} J. Akimiitsu, Symposium on Transition Metal Oxides, Sendai, January 10, 2001;
J. Nagamatsu, N. Nakagawa, T. Muranaka, Y. Zenitani, and J. Akimitsu (to be published).
\bibitem{budko}  S. L. Bud'ko, G. Lapertot, C. Petrovic, C. E. Cunningham,
N. Anderson, and P. C. Canfield, Phys. Rev. Lett. {\bf 86}, 1877 (2001).
\bibitem{DKF} D. K. Finnemore, J. E. Ostenson, S. L. Bud'ko, 
G. Lapertot, and P. C. Canfield, cond-mat/0102114.
\bibitem{canfield} P. C. Canfield, D. K. Finnemore, S. L. Bud'ko, J. E. Ostenson, 
G. Lapertot, C. E. Cunningham, and C. Petrovic, cond-mat/0102289.
\bibitem{jap1} Y. Takano, H. Takeya, H. Fujii, T. Hatano, K. Togano, H. Kito, 
and H. Ihara, cond-mat/0102167.
\bibitem{wisc} D. C. Larbalestier, M. Rikel, L. D. Cooley, A. A. Polyanskii, J. Y. Jiang, 
S. Patnaik, X. Y. Cai, D. M. Feldmann, A. Gurevich, A. A. Squitier, M. T. 
Naus, C. B. Eom, E. E. Hellstrom, R. J. Cava, K. A. Regan, N. Rogado, M. A. 
Hayward, T. He, J. S. Slusky, P. Khalifah, K. Inumaru, and M. Hass, 
cond-mat/0102216.
\bibitem{caplan} Y. Bugoslavsky, G. K. Perkins, X. Qi, L. F. Cohen, and
A. D. Caplin, cond-mat/0102353.
\bibitem{WHH} N. R. Werthamer, E. Helfand, and P. C. Hohenberg, Phys. Rev.
{\bf 147}, 295 (1966) and refs. therein.
\bibitem{boro1} K. D. D. Rathnayaka, A. K. Bhatnagar, A. Parasiris,
D. G. Naugle, P. C. Canfield, and B. K. Cho, Phys. Rev. B {\bf 55}, 8506 (1997).
\bibitem{boro2} V. Metlushko, U. Welp, A. Koshelev, I. Aranson, G. W. Crabtree, and
P. C. Canfield, Phys. Rev. Lett. {\bf 79}, 1738 (1997).
\bibitem{boro3} S. V. Shulga,
S.-L. Drechsler, G. Fuchs, K.-H. Muller, K. Winzer, M. Heinecke, and
K. Krug, Phys. Rev. Lett. {\bf 80}, 1730 (1998).
\bibitem{pippard} see for example: A. B. Pippard, Magnetoresistance in metals (Cambridge 
University Press, Cambridge, England, 1989).
\bibitem{korea} C. U. Jung, Min-Seok Park, W. N. Kang, Mun-Seog Kim, S. Y. Lee, and
Sung-Ik Lee, cond-mat/0102215.
\bibitem{dia1} S. L. Bud'ko, P. C. Canfield, C. H. Mielke, and A. H. Lacerda,
Phys. Rev. B {\bf 57}, 13624 (1998).
\bibitem{dia2} K. D. Myers, S. L. Bud'ko, I. R. Fisher,
Z. Islam, H. Kleinke, A. H. Lacerda, and P. C. Canfield, J. Magn. Magn.
Mater., {\bf 205}, 27 (1999).

\end{references}
\end{document}